\documentclass{article}
\usepackage{amsmath}
\usepackage{appendix}
\usepackage[retainorgcmds]{IEEEtrantools}
\usepackage{graphicx}
\usepackage{xcolor}
\usepackage{enotez}
\usepackage[super,sort&compress]{natbib}
\let\footnote=\endnote
\begin{document}
\title{Quantum corrections and multioccupancy in a semi-classical gas}
\author{Loris Ferrari \\ Department of Physics and Astronomy (DIFA) of the University of Bologna\\via Irnerio, 46 - 40126, Bologna,Italy}
\maketitle
\begin{abstract}
The quantum corrections to the behavior of a semi-classical gas can be expressed as power series of the ratio $\eta$ between the cube of De Broglie's thermal wavelength and the specific volume. The connection between $\eta$ and the multioccupancy of quantum states is the aim of the present work. By means of a chemical/physical approach, it is possible to associate $\eta$ to the concrete realization of multioccupancy in momentum space, through the formation of what we call \textquoteleft pseudo-molecules\textquoteright, i.e. multiplets of particles sharing momenta whose differences are negligible in the continuous spectrum approximation. The pseudo-molecules result as casual consequences of multiple scattering processes, among the non interacting (scattering apart) particles. \newline
\newline
\textbf{PACS:}   \newline 
\textbf{Key words:} Quantum gases; Semi-classical limit; Molecular gases; 
\end{abstract}

e-mail: loris.ferrari@unibo.it

\section{Introduction}
\label{Intro}

Ideal gases, formed by non interacting, point-like particles, show macroscopic quantum effects, like the Bose-Einstein condensation of Bosons and the stacking of Fermions below the Fermi level. Both phenomena share a common origin: the reduction of the number of states available for the particles, at low temperatures and/or high pressures. Those large quantum effects are thereby associated to an increasing \textquoteleft crowding\textquoteright$\:$of the quantum spectrum, i.e. to an increasing probability of \emph{multioccupancy} of the quantum states.

The present work is focused on the relationship between quantum effects and multioccupancy, in a gas of $N$ point-like particles of mass $m$, in a volume $V$, at high temperature $T$ and/or low pressure. In this semi-classical case, the (small) quantum corrections can be expressed as power series of the quantity:

\begin{subequations}
\label{eta,P}

\begin{equation}
\label{eta}
\eta:=\frac{N\mathrm{h}^3}{V(2\pi m k_BT)^{3/2}}<<1
\end{equation}
\\
($\mathrm{h}=$ Planck constant, $k_B=$ Boltzmann constant), as outlined in Section \ref{Pre}. The pressure, for instance, reads:\cite{TKS}

\begin{equation}
\label{P}
P=\frac{N k_BT}{V}\left[1\pm\frac{1}{2^{5/2}g_A}\eta+\mathbf{o}(\eta^2)\right]\:,
\end{equation}
\end{subequations}
\\
for Fermions ($+$) and Bosons ($-$), where $g_A$ is the spin degeneracy (see Subsection \ref{Spin}), and $\mathbf{o}(\cdot)$ indicates a series whose dominant term is linearly small in the argument. 

In the pedagogical literature, multioccupancy of a given energy level is sometimes considered in a probabilistic context.\cite{Pathria} It is also stressed that $\eta$ is the ratio between the thermal de Broglie's wavelength and the specific volume $V/N$. More than this, Section \ref{Pseudo} elucidates the connection between $\eta$ and multioccupancy, through a chemical procedure. It is shown that $\eta^{j}$ is proportional to the concentration of a special kind of molecules, that we call \textquoteleft pseudo-molecules\textquoteright, formed by $j+1$ particles, sharing the same momentum, in the case of spinless Bosons, for a time comparable to the mean free time between two scattering events (Subsection \ref{PG}). 

In order to include the spins (in particular, when Pauli's exclusion principle forbids Fermions' multioccupancy), the notion of pseudo-molecule is extended to particles whose momentum differences do exist, but are negligible in the continuous spectrum approximation (Subsection \ref{Spin}). The results are the same as in the spinless Bosons' case.

In conclusion, independently from the spinorial nature of the gas, the $j$-th term of any expansion like Eqn \eqref{P} turns out to be proportional to the concentration of \textquoteleft pseudo-polymers\textquoteright, formed - for pure statistical reasons - by $j+1$ particles. This result strengthens the tight relationship between quantum corrections and multioccupancy in semi-classical gases, and drags to the light the quantum core nested in the perfect gases, a subject that might be obscured by the absence of Planck's constant $\mathrm{h}$ in the expressions of measurable quantities, like heat, pressure and entropy's differences. However, $\mathrm{h}$ enters explicitly in the expression of the chemical potential $\mu=KT\ln \eta$, and in the (measurable) equilibrium concentrations of aggregates of particles, undergoing dissociation and recombination processes, even though the expressions used for the calculations refer just to perfect gases. A paradigmatic case is Saha formula, i.e. the concentration of free electrons in a semi-classical gas of Hydrogen.\cite{Nickel} The pseudo-molecules provide another example of purely statistical dissociation/recombination processes extracting $\mathrm{h}$ from the quantum core of the perfect gas.     

\section{Quantum corrections in semi-classical gases}
\label{Pre}

The full theory of ideal quantum gases, for Bosons and Fermions\cite{TKS,Linder,Swendsen,Carter,Pathria2,Bowley,Huang} shows the crucial role played by the chemical potential $\mu$, satisfying the equation:

\begin{equation}
\label{FB}
1=\frac{1}{N}\int_0^\infty\frac{g(\epsilon)}{\mathrm{e}^{(\beta\epsilon-\mu)}\pm1}\mathrm{d}\epsilon\:,
\end{equation}
\\
where $\beta=1/(k_B T)$, and 

\begin{equation}
\label{g,n(epsilon)}
g(\epsilon)=\frac{V2\pi(2m)^{3/2}g_A}{\mathrm{h}^3}\sqrt{\epsilon}
\end{equation}
\\
is the free-particle density of states (in 3 spatial dimensions). $+$/$-$ refer to Fermions/Bosons respectively. In particular, it is seen that the vanishing of $\mathrm{e}^{\beta\mu}$ recovers the perfect gas thermodynamics. The condition $\mathrm{e}^{\beta\mu}<<1$ implies:\cite{Bowley,Huang,Roy}  

\begin{equation}
\label{eta2}
\mathrm{e}^{\beta\mu}\rightarrow\eta:=\frac{N\mathrm{h}^3}{V(2\pi mK T)^{3/2}}=\frac{N}{Z_0(m)}\:,
\end{equation}
\\
where

\begin{equation}
\label{Zcm}
Z_0(m):=\int_0^\infty\mathrm{d}\epsilon g(\epsilon)\mathrm{e}^{-\beta\epsilon}=\frac{V(2m\pi k_BT)^{3/2}}{\mathrm{h}^3}
\end{equation}
\\
is the single-particle partition function. Explicitating the dependence on the mass $m$ in $Z_0(m)$ will be useful in what follows. 

The quantity $\eta$ is assumed as the parameter which keeps quantum effects under control, and makes Eqn \eqref{eta} a necessary condition for the semi-classical limit. Actually, a series expansion in powers of $\mathrm{e}^{\beta\mu}$ of the right-hand side of Eqn \eqref{FB} is the first step of a mathematical procedure leading to expressions like \eqref{P}, for any thermodynamical quantity of interest.

\section{Quantum effects and pseudo-molecules}
\label{Pseudo}  

In a standard diatomic molecule (dimer) $A_2$, two atoms $A$ (identical, for the present aims) form a pair, characterized by a positive dissociation energy $\epsilon_{diss}$, i.e. the energy required to break the covalent (for identical atoms) interatomic bond. For $k_BT<< \epsilon_{diss}$, the molecule is stable, against the typical impact energy due to the thermal motions. Hence, the molecular lifetime is very long, compared to the mean time $\tau$ between two scattering events. During this relatively long time, the molecule $A_2$ behaves like a bond structure, whose centre of mass moves freely in space. 

Loosely speaking, the simplest form of what we call pseudo-molecule is nothing but a dimer with $\epsilon_{diss}=0$, i.e. with a vanishing dissociation energy.\cite{GiantMol} At a first sight, one might be tempted to conclude that the pseudo-molecules are simply nonsensical, but this conclusion ignores the effects of statistics: even if any scattering event can dissociate a pseudo-molecule, no matter how small the impact energy, any scattering event has, in turn, a non vanishing probability of \emph{creating} one, in the form of two particles emerging with the same momentum. In contrast to a \textquoteleft true\textquoteright$\:$ dimer, the lifetime of such pseudo-dimer is very short, and comparable to the mean free time $\tau$ between two scattering events. What looks interesting is that the equilibrium concentration of the pseudo-molecules can be calculated with the standard methods of molecular chemistry. Actually, the concentration of \emph{any} possible pair of particles, even with different momenta, can be calculated accordingly, as shown in Subsection \ref{Spin}.

\subsection{Pseudo-molecules in a perfect gas of spinless Bosons}
\label{PG}
 
The minimization of the Helmholtz free energy $F(T,\:V,\:X)$ with respect to any parameter $X$, at constant temperature $T$ and volume $V$, determines the equilibrium value $X(T,\:V)$. This is the method we use here to calculate a molecular concentration, in alternative to the mass action formula,\cite{Carrington} used in chemical physics. For a perfect gas, formed by point-like spinless Bosons, one has (recall Eqn \eqref{Zcm}):\cite{RT-K}

\begin{align}
F(T,\:V,\:N)&=-k_BTN\ln\Big[\frac{V(2m\pi k_BT)^{3/2}\mathrm{e}}{N\mathrm{h}^3}\Big]=\nonumber\\
\label{phiPG}\\
&=-k_BTN\ln\Big[\frac{Z_0(m)\mathrm{e}}{N}\Big]\nonumber\:,
\end{align}
\\
where \textquoteleft $\mathrm{e}$\textquoteright$\:$ results from $N! \approx (N/\mathrm{e})^N$ (a consequence of Stirling approximation \cite{Stir}).

A doubly occupied \emph{classical} state would be a pair of point-like particles $A$ sharing the same position $\mathbf{r}$ and momentum $\mathbf{p}$.  Hence, a layman picture, suitable for undergraduate students, suggests to identify a doubly occupied state with a point-like particle $A_2$, with doubled mass and the same momentum. The chemical reaction accounting for the formation/dissociation of such \textquoteleft pseudo-molecule\textquoteright$\:$ is necessarily a many-body scattering event:  

\begin{equation}
\label{Reac}
A+A+A+\cdots\longleftrightarrow A_2+A+\cdots\:,
\end{equation} 
\\
in order that energy and momentum are conserved. In thermal equilibrium there will be $N_2$ pseudo-molecules and $N-2N_2$ single particles. Hence, the total Helmholtz free energy of the gas will be the sum of the free energies of the two gases (single particles and pseudo-molecules). From Eqns \eqref{phiPG}, one gets:

\begin{subequations}
\label{phitot}
\begin{align}
F_{tot}(T,\:V,\:N_2)&=\overbrace{-k_BTN_2\ln\Big[\frac{Z_0(2m)\mathrm{e}}{N_2}\Big]}^{F_{A_2}(T,\:V,\:N_2)}-\label{phiA2}\\
\nonumber\\
&\underbrace{-k_BT(N-2N_2)\ln\Big[\frac{Z_0(m)\mathrm{e}}{(N-2N_2)}\Big]}_{F_A(T,\:V,\:N-2N_2)}\label{phiA}\:.
\end{align}
\end{subequations}
\\
Notice the total mass $2m$ in the line \eqref{phiA2}. The equilibrium value of the fraction of pseudo-molecules $N_2/N$ simply follows from the equation $\partial F_{tot}/\partial N_2=0$, which yields, from Eqns \eqref{phitot} and \eqref{Zcm}:

\begin{equation}
\label{rpg}
\frac{N_2}{N}=\frac{Z_0(2m)N}{Z_0(m)^2}\left[1-2N_2/N\right]^2=\frac{N\mathrm{h}^3}{V(\pi mk_BT)^{3/2}}\left[1-2N_2/N\right]^2\:.
\end{equation}
\\
Finally, from Eqns \eqref{rpg} and \eqref{eta}, one gets:

\begin{equation}
\label{P1N2/N}
\eta=\frac{1}{2^{3/2}}\frac{N_2}{N}[1+\mathbf{o}(N_2/N)]\:,
\end{equation}
\\
which shows that the semi-classical condition \eqref{eta} simply requires the smallness of the pseudo-molecules' concentration $N_2/N$ in thermal equilibrium. 

Using the same method of minimizing the Helmohltz free energy, It is easy to show that the equilibrium concentration $N_j/N$ of  \textquoteleft pseudo-polymers\textquoteright$\:$$A_{j}$, formed by $j$ Bosons, moving with the same momentum reads:

\begin{align}
\label{Nlambda}
\frac{N_j}{N}&=\frac{Z_0(j m)N^{j-1}}{Z_0(m)^j}\left[1-j\frac{N_j}{N}\right]^j\nonumber\\
\nonumber\\
&=j^{3/2}\left[\frac{N}{Z_0(m)}\right]^{j-1}\left[1-j\frac{N_j}{N}\right]^j=j^{3/2}\eta^{j-1}[1+\mathbf{o}(\eta)]\:.
\end{align}
\\
Equation \eqref{Nlambda} shows that $\eta^j\propto N_{j+1}/N$. Hence, the $j$-th order term of any series expansions like Eqn \eqref{P} is proportional to the concentration of pseudo-molecules $A_{j+1}$, formed by $j+1$ particles. This concludes the complete identification of the quantum corrections with multioccupancy (for spinless Bosons).

\subsection{Including the spin}
\label{Spin} 

Even undergraduate students know (from a preliminary class of Chemistry, for instance) that two Fermions (electrons in particular) cannot occupy the same orbital, if their spins are both \textquoteleft up\textquoteright$\:$or  \textquoteleft down\textquoteright. It is possible, in principle, to orient almost all the spins up or down, by applying a strong magnetic field. For such polarized gas, the formation of Fermions' pairs sharing the same momentum would be extremely unlikely. So, the presence of the spin seems to deny the tight (and universal) relationship between quantum corrections and multioccupancy, as deduced for spinless particles. What we are going to show, is that the spin has no practical influence in the continuous spectrum approximation, as a consequence of the thermodynamic limit

\begin{equation}
\label{TL}
V,\:N\rightarrow\infty\quad;\quad \frac{N}{V}<\infty\:.
\end{equation}
\\  

Let us consider an \emph{unpolarized} gas of spinors $A$, in the absence of fields favoring some spin direction(s) with respect to others. The spinor partition function simply accounts for the internal degeneracy $g_A=2s_A+1$, due to the possible different projections of the spin along a given axis:

\begin{equation}
\label{ZA}
Z_A=g_AZ_0(m)\:.
\end{equation} 
\\
If $s_A\geq0$ is integer ($g_A$ odd), the spinor is a Boson; if $s_A>0$ is half integer ($g_A$ even), the spinor is a Fermion. Generalizing equation \eqref{P1N2/N}, due to the presence of $g_A$ in Eqn \eqref{ZA}, one gets:

 \begin{equation}
 \label{etasp}
 \eta_{sp}=\frac{N}{Z_{A}}=\frac{N_2/N}{g_A2^{3/2}}[1+\mathbf{o}(N_2/N)]\:,
 \end{equation}
 \\
for the parameter controlling the quantum corrections in the spinorial case. The spin of dimers formed by identical atoms results from the sum of the atomic spins and yields the degenercy $g_A^2=g_++g_-$, where $g_+$ is the number of states symmetric for the spins' exchange (\emph{even} spin parity), and $g_-$ is the number of antisymmetric states (\emph{odd} spin parity).   

Let us extend the notion of pseudo-molecule to \emph{any} couple of free particles (Fermions or Bosons), emerging from a scattering like Eqn \eqref{Reac}, with \emph{different} momenta $\vec{p}\pm\Delta\vec{p}$. Since the two particles of mass $m$ are free after the scattering, the pair can be regarded as to a \textquoteleft dimer\textquoteright, with mass $2m$ and momentum $2\vec{p}$, whose \textquoteleft internal\textquoteright$\:$state is represented by two masses moving in opposite directions, with momenta $\pm\Delta\vec{p}$, in the reference system of the pseudo-dimer's centre of mass. Hence, the internal energy spectrum of the pseudo-dimer is nothing but the kinetic energy $\Delta \vec{p}^2/m$, that is independent from the $g_A^2$ states, resulting from the sum of the particles' spins. The partition function is, by definition, the sum of the Boltzmann factors of the energy eigenvalues, times the corresponding degeneracies. The internal partition function then reads:

\begin{subequations}
\label{zintZ2spin}
\begin{equation}
\label{zint}
z_{int}=g_A^2\mathrm{e}^{-\beta\Delta \vec{p}^2/m}\:,
\end{equation}
\\      
and the partition function of the whole pair reads, in turn:

\begin{equation}
\label{ZA2spin}
Z_{A_2}=Z_0(2m)z_{int}=g_A^2Z_0(2m)\mathrm{e}^{-\beta\Delta \vec{p}^2/m}\:,
\end{equation}
\end{subequations}
\\
including the contribution $Z_0(2m)$ from the pseudo-dimer centre of mass. The total Helhmoltz free energy of spinors gas, containing $N_2(\Delta p)$ pairs, is similar to Eqns \eqref{phitot}, apart from the substitutions $Z_0(2m)\rightarrow Z_{A_2}$ (Eqn \eqref{ZA2spin}) inside the square brakets of line \eqref{phiA2} and $Z_0(m)\rightarrow Z_A$ (Eqn \eqref{ZA}) inside the square brakets of line \eqref{phiA}. The concentration of pseudo-dimers with different momenta follows from the Helmholtz free energy's minimization: 

\begin{equation}
\label{NDelta/N}
\frac{N_2(\Delta\vec{p})}{N}=\frac{N_2}{N}\mathrm{e}^{-\beta\Delta\vec{p}^2/m}=\frac{N_2}{N}[1+\mathbf{o}(\beta\Delta\vec{p}^2/m)]\:,
\end{equation}
\\
and is \emph{spin independent}, until $\Delta\vec{p}\ne0$. However, the spatial parity of the two-particle state (symmetry/antisymmetry for co-ordinates' exchange) forbids the occupancy of the same momentum for the $g_+$ fermionic states with \emph{even} spin parity, and for the $g_-$ bosonic states with \emph{odd} spin parity.\cite{Parity} Hence, if $\Delta\vec{p}=0$, the pseudo-dimer formation implies the substitutions $g_A^2\rightarrow g_-$ for Fermions and $g_A^2\rightarrow g_+$ for Bosons in Eqns \eqref{zintZ2spin}. The resulting effect is $N_2(0)/N=(g_\mp/g_A^2)N_2/N$ for Fermions ($-$) and Bosons ($+$). This is the reason why the spins do have an explicit influence on the concentration of pseudo-dimers with $\Delta\vec{p}=0$. The final step is showing that this is irrelevant, in the thermodynamic limit. 

As it is well known from elementary quantum mechanics, in a cubic (for simplicity) volume $V=L^3$, the momentum eigenvalues are $\vec{p}=(\hbar/L)\vec{n}$, $\vec{n}$ being a vector with integer Cartesian components. Hence the modulus of any difference between two momenta reads $2|\Delta\vec{p}|=J\hbar/L$, with $J$ a natural number, and vanishes in the thermodynamic limit $L\rightarrow\infty$, unless $J\propto L$. This argument is just what underlies the continuous spectrum approximation, in which the difference between nearest neighbor energy levels is much smaller than the thermal energy $k_BT$, and the energy spectrum can be treated as continuous. \cite{Landsberg} Looking at the second equality in Eqn \eqref{NDelta/N}, it is seen that the double (or multiple) occupancy expressed by $N_2/N$ (or $N_j/N$) does not refer only to particles sharing the \emph{same} momentum, but to particles whose \emph{different} momenta are close enough to satisfy the continuous spectrum approximation, i.e. the condition:

\begin{equation}
\label{CSA2}
\beta\Delta\vec{p}^2/m=\frac{(J\hbar)^2}{mL^2k_BT}\rightarrow0\quad;\quad J/L\rightarrow0
\end{equation}
\\
for $L\rightarrow\infty$. Hence, the paramenters $\eta$ or $\eta_{sp}$ (Eqn \eqref{etasp}), and their (finite) powers, can be identified with the multioccupancy in \emph{momentum} space, of energy layers, whose width vanishes in the thermodynamic limit, regardless to the spinorial nature of the gas particles.

\section{Conclusions}
\label{Concl}

The evidence that quantum effects in gases increase with increasing \textquoteleft crowding\textquoteright of the spectrum (as shown by Bose-Eistein condensation and by the stacking of Fermions below the Fermi level), suggests a possible relationship between multioccupancy of states and quantum effects. The question can be investigated in the realm of semi-classical gases, by using a  
chemical approach (Section \ref{Pseudo}). 

In Subsection \ref{PG} the notion of \textquoteleft pseudo-molecule\textquoteright$\:$ is introduced for spinless Bosons, first as a pair, then as a $j$-ple of particles, casually sharing the same momentum, for a time comparable to the mean free time. On applying the standard methods of chemical-physics, it is shown that the (small) equilibrium concentration of the pseudo-molecules is proportional to the $(j-1)$-th power of the ratio $\eta$ between the cube of de Broglie's thermal wavelength and the specific volume. Since the quantum corrections in a semi-classical gas are expressed as series expansion in powers of $\eta$, the $j$-th order term of such series is proportional to the concentration of pseudo-polymers, formed by $j+1$ spinless Bosons. This result can be generalized to spinors too (Subsection \ref{Spin}), provided the notion of pseudo-molecules does include particles occupying momenta close enough to one another, as to satisfy the continuous spectrum approximation. It is thereby seen that the quantum corrections are caused by the multioccupancy in \emph{momentum space}, realized through the formation of what we call pseudo-molecules.

In a perfect gas, the expressions of measurable thermodynamical quantities like heat, pressure, entropy differences, do not contain the Planck constant $\mathrm{h}$ at all. However, $\mathrm{h}$ explicitly emerges in the \emph{measurable} concentration of chemical species resulting from dissociation/recombination prcesses, even if the expressions used refer to \emph{perfect} gases in thermal equilibrium.\cite{Carrington, Nickel} The dissociation/recombination of the pseudo-molecules provides a further example of appearence of $\mathrm{h}$, showing that the quantum core of the perfect gases is nested in the multioccupancy in momentum space.
\\
\underline{Statement}: The author has no conflict of interest to declare

\printendnotes

\begin{thebibliography}{20}

\bibitem{TKS} M. Toda, R. Kubo and N. Saito, \emph{Statistical Phisics I: Equilibrium Statistical Mechanics}, Springer-Verlag - \textbf{Berlin, Heidelberg, New York, London, Paris, Tokyo, Hong Kong, Barcelona, Budapest}, 86-88, (1978).

\bibitem{Pathria} See, in particular: R.K. Pathria and P.D. Beale, \emph{Statistical Mechanics}, Elsevier - \textbf{Amsterdam, Boston, Heidelberg, Londin, New York, Oxford, Paris, San Diego, San Francisco, Siingapore, Sidney, Tokyo}, p 142 (2011).

\bibitem{Nickel} G.H. Nickel, \emph{Elementary derivation of Saha equation}, Am. J. of Phys. \textbf{48}, 448-450 (1980).

\bibitem{Linder} B. Linder, \emph{Thermodynamics and introductory Statistical Mechanics} Part II, Wiley-Interscience  - \textbf{Hobokn, New Jersey}, 127-143 (2004).

\bibitem{Swendsen} R.H. Swendsen, \emph{An introduction to Statistical Mechanics and Thermodynamics}, Oxford University Press, 350-385 (2020).

\bibitem{Carter} A.H. Carter, \emph{Classical and Statistical Thermodynamics}, Prentice Hall - \textbf{Upper Saddle River, New Jersey}, 215-259 (2001). 

\bibitem{Pathria2} See ref [2], pp 143-155.

\bibitem{Bowley} R. Bowley, and M. Sanchez, \emph{Introductory Statistical Mechanics}, Oxford Science Publications, Clarendon Press - \textbf{Oxford}, 211- 218 (1999).

\bibitem{Huang} Huang. K, \emph{Statistical mechanics}, Wiley - \textbf{New York, Chichester, Brisbane, Toronto, Singapore}, 179-189 (1987).

\bibitem{Roy} B.N. Roy, \emph{Fundamentals of Classical and Statistical Thermodynamics}, John Wiley \& Sons, 297-306 (2002).

\bibitem{GiantMol} Examples of dimers with very small dissociation energy are the \emph{giant} molecules, formed by He atoms at very low temperatures. See J. Leonard, A.P. Mosk, M. Walhout, P. van der Straten, M. Leduc and C. Cohen-Tannoudji, \emph{Analysis of Photoassociation Spectra for Giant Helium Molecules}, Phys. Rev. A  \textbf{69}, 032702, 1-12 (2004).

\bibitem{Carrington} G. Carrington, \emph{Basic Thermodynamics}, Oxford Science Publicaions - \textbf{Oxford, New York, Tokyo}, 332-337 (1995).

\bibitem{RT-K} See ref [1], p 54; ref [6], p 150, and R. Tahir-Kheli: \emph{General and statistical thermodynamics}, Springer (Heidelberg, Dodrecht, London, New York) (2012), p 415. 

\bibitem{Stir} M.R. Spiegel, \emph{Complex Variables}, Schaum's Outline Series, McGraw-Hill Book Company -  \textbf{New York, St. Louis, San Francisco, Toronto, Sydney}, 275 (1964).

\bibitem{Parity} By principle, the \emph{total} parity of a two-particle quantum state must be \emph{odd} for Fermions and \emph{even} for Bosons. Since in the present case [total parity] = [spatial parity]$\times$[spin parity], spatial and spin parities must be opposite for Fermions, and identical for Bosons. The odd spatial parity - which forbids double occupancy of the same momentum - applies to the $g_+$ fermionic states with even spin parity, and to the $g_-$ bosonic states with odd spin parity. Hence the pseudo-dimer's spin degeneracy reduces to $g_-$ for Fermions and $g_+$ for Bosons, if $\Delta\vec{p}=0$.     
 
\bibitem{Landsberg} P.T. Landsberg, \emph{The continuous spectrum approximation in quantum statistics}, Phys. Rev. \textbf{94}, 469-471 (1954).

\end{thebibliography}
\end{document}